\documentstyle[preprint,eqsecnum,aps,epsf]{revtex}

\tightenlines
\begin{document}
\preprint{\vbox{
\hbox{INPP-UVA-98-03} 
\hbox{January, 1998} 
\hbox{hep-ph/9804461} 
}}
\draft
\def\be{\begin{eqnarray}}
\def\en{\end{eqnarray}}
\def\non{\nonumber}
\def\la{\langle}
\def\ra{\rangle}
\def\up{\uparrow}
\def\dw{\downarrow}
\def\ep{\varepsilon}
\def\ms{\overline{\rm MS}}
\def\ums{{\mu}_{_{\overline{\rm MS}}}}
\def\u{\mu_{\rm fact}}
\def\pr{{\sl Phys. Rev.}~}
\def\ijmp{{\sl Int. J. Mod. Phys.}~}
\def\mpl{{\sl Mod. Phys. Lett.}~}
\def\prp{{\sl Phys. Rep.}~}
\def\prl{{\sl Phys. Rev. Lett.}~}
\def\pl{{\sl Phys. Lett.}~}
\def\np{{\sl Nucl. Phys.}~}
\def\ppnp{{\sl Prog. Part. Nucl. Phys.}~}
\def\zp{{\sl Z. Phys.}~}

\title{\bf Baryon Magnetic Moments and Quark Orbital Motion\\
 in\\
the Chiral Quark model\\}

\author{\bf Xiaotong Song
\\}

\address{Institute of Nuclear and Particle Physics\\
Jesse W. Beams Laboratory of Physics\\
Department of Physics, University of Virginia\\
Charlottesville, VA 22901, USA\\}
\date{January 5, 1998, revised April 29, 1998}
\maketitle
\vskip 12pt
\begin{abstract}
Using the unified scheme for describing both quark spin and 
orbital angular momenta in the chiral quark model developed 
in the previous work, the magnetic moments of octet and decuplet 
baryons are calculated. The numerical result shows that the 
overall agreement with data is improved by including the orbital 
contributions.
\end{abstract}
\bigskip
\bigskip
\bigskip


\newpage

\leftline{\bf I. Introduction}

Although the naive SU(6) constituent quark model (NQM) is attractively
simple, it has had only limited quantitative success in accounting for 
the magnetic moments and semileptonic decays of the baryons. In the 
NQM all three quarks in the nucleon are assumed to be in the s-wave 
state, the nucleon spin is completely attributed to the quark spin
and the orbital angular momentum (OAM) is zero $<L_z>_q=0$. The EMC 
\cite{emc} and recent experiments \cite{smc,slac,hermes} on deep 
inelastic lepton-nucleon scatterings (DIS) show that the quark spin 
in NQM cannot account for the proton spin and lead to `spin crisis'. 
In an earlier work \cite{sehgal}, by relating the quark spin fractions, 
nucleon magnetic moments and the weak axial coupling constants of 
baryons, Sehgal shown that a large portion of nucleon spin arises 
from the orbital motion of the constituents. This idea has been 
generalized to all octet baryons to explain both the DIS quark spin 
fractions and baryon magnetic moments \cite{karl,br,sg}. In these works,
the quark orbital contributions are not explicitly included. In the past
decade, the important role of the OAM in the nucleon has been discussed 
in different models and various forms, an incomplete list see 
\cite{rat87,bali,ji,cs97,cl2,song9708}. In our previous paper 
\cite{song9708}, a unified scheme for describing both quark spin and 
orbital angular momenta in the symmetry breaking chiral quark model 
has been suggested. The spin and orbital angular momenta carried by 
the quarks and antiquarks are evaluated. In this paper, we use the 
results given in \cite{song9708} to calculate the baryon magnetic 
moments and discuss the effects of the orbital contributions.

The magnetic moment of the baryon $B$ can be written as
$$\mu_{B}=\sum\limits_{q}\mu_q[(\Delta q)^{B}
-(\Delta\bar q)^{B}
+<L_z>^{B}_{q}-<L_z>^{B}_{\bar q}]
\eqno (1a)$$
where $\mu_q$s are the magnetic moments of quarks, $\Delta q\equiv 
q_{\up}-q_{\dw}$ and $\Delta\bar q\equiv \bar q_{\up}-\bar q_{\dw}$,
and the notation $B$ denotes the member of the baryon octet or
decuplet. $q_{\up,\dw}$ (${\bar q}_{\up,\dw}$) are quark (antiquark) 
{\it numbers} of spin parallel and antiparallel to the nucleon spin,  
or more precisely, quark numbers of {\it positive} and {\it negative}
helicities, if the proton helicity is chosen to be $+1/2$.
$<L_z>^{B}_{q,\bar q}$ denotes the total orbital angular momentum 
carried by {\it quarks and antiquarks} in the baryon $B$. In the 
zeroth order approximation of the chiral quark model, the baryon is 
assumed to contain only three valence quarks. The antiquarks are 
produced from the first order chiral splitting processes 
$q\to q'+{\rm GB}(q_i\bar q_j)$, where GB denotes the Goldstone boson. 
We assume that the magnetic moment of the baryon is the sum of spin 
and orbital magnetic moments of individual charged particles (quarks 
or antiquarks). The assumption of {\it additivity} is commonly 
believed to be a good approximation for a loosely bound composite 
system, which is the basic description for the baryon in the 
effective chiral quark model. In reality the baryon may contains 
other neutral particles, such as gluons, as pointed out in 
\cite{bs90}. Although the gluons do not make any contribution to 
the magnetic moment in (1a), the existence of intrinsic gluons would 
significantly change the valence quark structure of the baryon due to 
the spin and color couplings between the gluon and quarks. For example, 
the total color charge for the gluon-quark system must be zero, and 
the total angular momentum must be 1/2. We will discuss a hybrid 
quark-gluon mixing model of the octet baryon in Sec. IV.

Since in the chiral quark model all antiquark sea polarizations are 
zero, $\Delta\bar q^{B}=0$ (the experiment \cite{smc96} shown that 
the antiquark sea polarization is rather small), hence
Eq.(1a) can be approximately can be rewritten as
$$\mu_{B}=\sum\limits_{q=u,d,s}\mu_q[(\Delta q)^{B}
+<L_z>^{B}_{q}-<L_z>^{B}_{\bar q}]
\eqno (1b)$$
where only three flavors of quark, $u$, $d$, and $s$ are considered.
To calculate the baryon magnetic moments (1b), we need to know not 
only the spin contents $\Delta q$, but also the orbital contents 
$<L_z>^{B}_{q}$ and $<L_z>^{B}_{\bar q}$ for all active quark 
(antiquark) flavors. All these quantities have been obtained in 
\cite{song9708}. For the purpose of later use, we briefly review
several key points and main results (Section II). The detail 
discussion can be found in \cite{song9708}. 

\bigskip

\leftline{\bf II. Spin and Orbital Motions in Chiral Quark Model.}

\leftline{\quad\bf A. Chiral quark model.}

The basic assumptions of the chiral quark model we used are: 
(i) the nucleon {\it flavor, spin} and {\it orbital} contents are
determined by its valence quark structure and all possible chiral 
fluctuations $q\to q'+{\rm GB}$, (ii) the probabilities of the chiral
splittings are rather small, one can treat the fluctuation 
$q\to q'+{\rm GB}$ as a small perturbation, for instance, the 
probability for the splitting $u\to d+\pi^-,~{\rm or}~d\to u+\pi^+$
is about $0.10-0.15$. The contributions from the higher order 
fluctuations can be neglected ($a^2<<1$). 

The effective Lagrangian describing interaction between quarks and 
the octet Goldstone bosons and singlet $\eta'$ is 

$${\it L}_I=g_8{\bar q}\pmatrix{({\rm GB})_+^0
& {\pi}^+ & {\sqrt\epsilon}K^+\cr 
{\pi}^-& ({\rm GB})_-^0
& {\sqrt\epsilon}K^0\cr
{\sqrt\epsilon}K^-& {\sqrt\epsilon}{\bar K}^0
&({\rm GB})_s^0 
\cr }q, 
\eqno (2a)$$
where $({\rm GB})_{\pm}^0=\pm {\pi^0}/{\sqrt 2}+
{\sqrt{\epsilon_{\eta}}}{\eta^0}/{\sqrt 6}+
{\zeta'}{\eta'^0}/{\sqrt 3}$, $({\rm GB})_s^0=
-{\sqrt{\epsilon_{\eta}}}{\eta^0}/{\sqrt 6}+
{\zeta'}{\eta'^0}/{\sqrt 3}$, and the symmetry 
breakings are explicitly included. The transition probability 
of chiral splitting $u(d)\to d(u)+\pi^{+(-)}$ is 
$a\equiv|g_8|^2$, and $\epsilon a$ denotes the probability of 
$u(d)\to s+K^{-(0)}$. Similar definitions are used for 
$\epsilon_\eta a$ and $\zeta'^2a$. Considering the mass 
suppression effect, one expects $0\leq\zeta'^2\leq 1$,
$0\leq\epsilon_{\eta}\leq 1$, and $0\leq\epsilon\leq 1$. 

The important feature of the chiral fluctuation is that due to the 
coupling between the quarks and GB's, a quark {\it flips its spin 
or helicity} and changes (or maintains) its flavor by emitting a 
charged (or neutral) Goldstone bosons. Therefore the flavor, spin, 
and orbital contents carried by the quark and antiquarks are
significantly different from those obtained without the chiral 
fluctuations.

For spin-up or spin-down valence $u$, $d$, and $s$ quarks, up to
the first order fluctuation, the allowed processes are
$$u_{\up,(\dw)}\to d_{\dw,(\up)}+\pi^+,~~
u_{\up,(\dw)}\to s_{\dw,(\up)}+K^+,~~
u_{\up,(\dw)}\to u_{\dw,(\up)}+({\rm GB})_+^0,~~
u_{\up,(\dw)}\to u_{\up,(\dw)}.
\eqno (3a)$$
$$d_{\up,(\dw)}\to u_{\dw,(\up)}+\pi^-,~~d_{\up,(\dw)}\to 
s_{\dw,(\up)}+K^{\rm 0},~~
d_{\up,(\dw)}\to d_{\dw,(\up)}+({\rm GB})_-^0,~~
d_{\up,(\dw)}\to d_{\up,(\dw)},
\eqno (3b)$$
$$s_{\up,(\dw)}\to u_{\dw,(\up)}+K^-,~~
s_{\up,(\dw)}\to d_{\dw,(\up)}+{\bar K}^{\rm 0},~~
s_{\up,(\dw)}\to s_{\dw,(\up)}+({\rm GB})_s^{\rm 0},~~
s_{\up,(\dw)}\to s_{\up,(\dw)}.
\eqno (3c)$$
The quark helicity flips in the chiral splitting processes 
$q_{\up,(\dw)}\to q_{\dw,(\up)}$+GB, i.e. the first three processes 
in each of (3a), (3b), and (3c). The consequences are, (i) the total 
spin content carried by quarks and antiquarks would be {\it smaller 
than} that without considering the chiral splitting, (ii) most
importantly, since the {\it quark spin flips} (or {\it helicity sign 
changes}) {\it in the fluctuations with} GB {\it emission,} the quark 
spin component changes one unit of angular momentum, $(s_z)_f-(s_z)_i
=+1$ or $-1$, the angular momentum conservation requires the {\it same 
amount change} of the orbital angular momentum but with {\it opposite 
sign}, i.e. $(L_z)_f-(L_z)_i=-1$ or $+1$. This {\it induced orbital 
motion} distributes among the quarks and antiquarks after the splitting. 
These two points are intimately related. 

The following combinations of the parameters $\epsilon$, 
$\epsilon_{\eta}$, and $\zeta'$ are useful in our formalism,
$$A\equiv 1-\zeta'+{{1-{\sqrt\epsilon_{\eta}}}\over 2},
\qquad  B\equiv \zeta'-{\sqrt\epsilon_{\eta}}\qquad
C\equiv \zeta'+2{\sqrt\epsilon_{\eta}}
\eqno (4a)$$
$$f\equiv{1\over 2}+{{\epsilon_{\eta}}\over 6}+{{\zeta'^2}\over 3},~~~
f_s\equiv{{2\epsilon_{\eta}}\over 3}+{{\zeta'^2}\over 3}
\eqno (4b)$$
and
$$\xi_1\equiv 1+\epsilon+f,~~~~~~~~~~\xi_2\equiv 2\epsilon+f_s
\eqno (4c)$$
The special combinations $A$, $B$ and $C$ stem from the combinations
of the octet and singlet neutral bosons appeared in the effective
chiral Lagrangian, while $f$ and $f_s$ stand for the transition 
probabilities of the chiral splittings $u_{\up}(d_{\up})\to
u_{\dw}(d_{\dw})+({\rm GB})_{+(-)}^0$ and $s_{\up}\to s_{\dw}+({\rm
GB})_s^0$ respectively. It is easy to see that the total transition 
probability of the first three processes in (3a), or (3b) is $\xi_1a$, 
and the corresponding probability in (3c) is $\xi_2a$.
\bigskip

\leftline{\quad\bf B. Quark spin contents}

Denoting the valence quark {\it numbers} in the baryon as 
$n^{(v)}_B(u_{\up})$, $n^{(v)}_B(u_{\dw})$, $n^{(v)}_B(d_{\up})$, 
$n^{(v)}_B(d_{\dw})$, $n^{(v)}_B(s_{\up})$, and $n^{(v)}_B(s_{\dw})$, 
the spin-up and spin-down quark (or antiquark) contents in the baryon 
$B$, up to the first order fluctuation, are
$$n_B(q'_{\up,\dw}, {\rm or}\ {\bar q'}_{\up,\dw}) 
=\sum\limits_{q=u,d,s}\sum\limits_{h=\up,\dw}
n^{(v)}_B(q_h)P_{q_h}(q'_{\up,\dw}, {\rm or}\ {\bar q'}_{\up,\dw})
\eqno (5)$$
where $P_{q_{\up,\dw}}(q'_{\up,\dw})$ and $P_{q_{\up,\dw}}({\bar
q}'_{\up,\dw})$ are the probabilities of finding a quark $q'_{\up,\dw}$
or an antiquark ${\bar q}'_{\up,\dw}$ arise from all chiral fluctuations 
of a valence quark $q_{\up,\dw}$. The probabilities,
$P_{q_{\up,\dw}}(q'_{\up,\dw})$ and $P_{q_{\up,\dw}}({\bar
q}'_{\up,\dw})$, depend on the effective interaction Lagrangian (2).
They were given in Table I in \cite{song9708}.
The spin-weighted quark contents are
$$(\Delta q')^{B}=\sum\limits_q [n^{(v)}_B(q_{\up})-n^{(v)}_B(q_{\dw})]
[P_{q_{\up}}(q'_{\up})-P_{q_{\up}}(q'_{\dw})]
\eqno (6a)$$
while the spin-weighted antiquark contents are zero
$$(\Delta\bar q')^B=0,
\eqno (6b)$$
due to
$P_{q_{\dw}}(q'_{\up})=P_{q_{\up}}(q'_{\dw})$,
$P_{q_{\dw}}(q'_{\dw})=P_{q_{\up}}(q'_{\up})$, and
$P_{q_{\dw}}({\bar q}'_{\up})=P_{q_{\up}}({\bar q}'_{\dw})=
P_{q_{\dw}}({\bar q}'_{\dw})=P_{q_{\up}}({\bar q}'_{\up})$.
In general the probabilities 
$P_{q_{\up,\dw}}(q'_{\dw,\up},~{\bar q'}_{\up,\dw})$ may
vary with the baryons, because the suppression effects may be different 
in different baryons. For the sake of simplicity, we assume that they are
universal for all baryons. Hence the $B$-dependence of the spin-weighted 
quark contents appears only in the $B$-dependence of
$n^{(v)}_B(q_{\up,\dw})$. 
\bigskip

\leftline{\quad\bf C. Quark orbital angular momentum}

In the quark splitting process, the flip of the quark spin will induce 
an orbital angular momentum. We assume that the induced orbital motion 
is {\it equally shared by the quarks and antiquarks after splitting} 
and introduce a {\it partition factor} $k$. Assuming the Goldstone 
boson has a simple quark structure, i.e. each boson consists of a 
quark and an antiquark, one has two quarks and one antiquark (total 
number is {\it three}) after each splitting $q\to q'$+GB. Hence up 
to first order splitting, one has $k=1/3$. The last processes in (3a), 
(3b), and (3c), make no contributions to the orbital motion. Similar to 
the probabilities $P_{q_{\up,\dw}}(q'_{\up,\dw})$ and $P_{q_{\up,\dw}}
({\bar q}'_{\up,\dw})$, we define $<L_z>_{q'/q_{\up}}$ 
($<L_z>_{{\bar q'}/q_{\up}}$) as the OAM carried by a {\it specific} 
quark $q'$ (antiquark $\bar q'$), arises from a valence spin-up quark 
$q_{\up}$ fluctuates into all allowed final states except for no 
emission case. The quantities $<L_z>_{q'/q_{\up}}$ and 
$<L_z>_{\bar q'/q_{\up}}$ for $q=u,d,s$ also depend on the effective 
Lagrangian (2) and have been given in Table II in Ref. \cite{song9708}. 
Again, we assume that the probabilities $<L_z>_{{q'}/q_{\up,\dw}}$ and 
$<L_z>_{{\bar q'}/q_{\up,\dw}}$ are {\it universal} for all baryons. 

The difference between the orbital angular momentum carried by quark 
$q$ and that carried by corresponding antiquark $\bar q$, for example 
for $u$-quark, is
$$<L_z>_{u}^B-<L_z>_{\bar u}^B=\sum\limits_{q}
[n^{(v)}_B(q_{\up})-n^{(v)}_B(q_{\dw})][<L_z>_{u/q_{\up}}-
<L_z>_{{\bar u}/q_{\up}}]
\eqno (7)$$
similar equations hold for $d$-quark and $s$-quark, and corresponding 
antiquarks. Where $\sum$ summed over all valence quarks in the
baryon $B$. 

Eqs.(1b), (6a), (6b) and (7) are main formulae to calculate the baryon
magnetic moments. Since the quantities $P_{q_h}(q'_{\up,\dw}, 
{\rm or}\ {\bar q'}_{\up,\dw})$, and $<L_z>_{{q',{\rm or},~\bar
q'}/q_{\up,\dw}}$ are known (Tables I and II in \cite{song9708}) and
universal for all baryons, we only need to know the valence quark 
numbers $n^{(v)}_B(q_{\up,\dw})$ in a specific baryon $B$, and these
numbers depend on the models of baryon. For our purpose of showing the 
effects of the orbital angular momentum, we only consider two special 
cases of quark valence structure: (1) static $SU(3)\otimes SU(2)$ model 
(Sec. III), and (2) hybrid quark-gluon mixing model (Sec. IV). We note
that similar discussions on the baryon magnetic moments in the chiral 
quark model without considering the orbital contributions were given in
\cite{wb,los}. We also note that a different version of including the
orbital contribution in a simple SU(3) symmetry chiral quark model was
independently discussed in \cite{cl2}.  

\bigskip

\leftline{\bf III. $SU(3)_f\otimes SU(2)_s$ Valence Structure (Model I).}

\leftline{\quad\bf A. Octet baryons.}

Assuming the flavor-spin piece of the valence quark structure is 
$SU(3)_f\otimes SU(2)_s$, the valence quark numbers in the proton
are
$$n^{(v)}_p(u_{\up})={5\over 3},~~n^{(v)}_p(u_{\dw})={1\over 3},~~
n^{(v)}_p(d_{\up})={1\over 3}, ~~n^{(v)}_p(d_{\dw})={2\over 3},~~
n^{(v)}_p(s_{\up,\dw})=0,
\eqno (8)$$
where $n^{(v)}_p(s_{\up,\dw})=0$ is due to no valence strange quarks 
exist in the proton. The quark spin contents $\Delta q$ and the 
difference between the quark and antiquark orbital angular momenta 
$<L_z>_{u,d,s}-<L_z>_{\bar u,\bar d,\bar s}$ in the octet baryons are 
listed in Table I. Using (1b) and Table I, the magnetic moments of 
octet baryons can be obtained. We make some remarks before going to
the numerical results.

It is easy to verify that the magnetic moments of the octet baryon 
satisfy the following sum rules 
$$(4.70)~~~{\mu}_p-{\mu}_n={\mu}_{\Sigma^+}-{\mu}_{\Sigma^-}-
({\mu}_{\Xi^0}-{\mu}_{\Xi^-})~~~(4.22)
\eqno (9a)$$
$$(3.66)~~~-6{\mu}_{\Lambda}=-2({\mu}_p+{\mu}_n+{\mu}_{\Xi^0}+{\mu}_{\Xi^-})
+({\mu}_{\Sigma^+}-{\mu}_{\Sigma^-})~~~(3.34)
\eqno (9b)$$
$$(4.14)~~~{\mu}_p^2-{\mu}_n^2=({\mu}_{\Sigma^+}^2-{\mu}_{\Sigma^-}^2)
-({\mu}_{\Xi^0}^2-{\mu}_{\Xi^-}^2)~~~(3.56)
\eqno (9c)$$
$$(0.33)~~~{\mu}_p-{\mu}_{\Sigma^+}={3\over
5}({\mu}_{\Sigma^-}-{\mu}_{\Xi^-})
-({\mu}_n-{\mu}_{\Xi^0})~~~(0.31)
\eqno (9d)$$
where the values of the two sides taken from the data \cite{pdg96}
are shown in parentheses. The relations (9a) and (9b) were first given 
by Franklin in \cite{frank1}. One can also show that our nonlinear
sum rule (9c) is equivalent to Eq. (16) given in \cite{frank2}. The
relations (9a), (9b), and (9c) are not new and violated at about 
$10-15\%$ level. They have been discussed in many works, for instance 
\cite{karl,br,sg}. However, the new relation (9d) is very well
satisfied. All these sum rules (9a)-(9d) also hold in the simple quark
model. Our result shows that if the $SU(3)\otimes SU(2)$ valence quark 
structure is used, the chiral fluctuations cannot change these sum rules 
even the orbital angular momenta are included. Furthermore, we have shown 
in \cite{sg} that the sum rules (9a)-(9c) also hold for more general case.  

Explicitly, (9a)-(9d) can be written as
$${\mu}_p-{\mu}_n={5\over 3}\delta_1(\mu_u-\mu_d)
\eqno (10a)$$
$${\mu}_{\Lambda}=-a\epsilon(\mu_u+\mu_d)+\delta_2\mu_s
\eqno (10b)$$
$${\mu}_p^2-{\mu}_n^2={5\over 3}\delta_1(\mu_u-\mu_d)
[(\delta_1+2a)(\mu_u+\mu_d)-2a\epsilon\mu_s]
\eqno (10c)$$
$${\mu}_p-{\mu}_{\Sigma^+}=-{1\over 3}\delta_3(\mu_d-\mu_s)
\eqno (10d)$$
where $\delta_1$, $\delta_2$, and $\delta_3$ are defined as
$$\delta_1=1-a(2\xi'_1-\epsilon-2)
\eqno (11a)$$
$$\delta_2=1-2a(\xi'_1-\epsilon)
\eqno (11b)$$
$$\delta_3=1-a[(1-\epsilon)+r_d(2\xi'_1-2\epsilon-1)
-r_s(2\xi'_2-3\epsilon)]/(r_d-r_s)
\eqno (11c)$$
where $r_{d,s}\equiv\mu_{d,s}/\mu_u$. If there are no chiral 
fluctuations, then $a=0$, $\delta_{1,2,3}\to 1$, and (10a)-(10d)
reduce to the simple quark model results
$$
{\mu}_p-{\mu}_n={5\over 3}(\mu_u-\mu_d),~~~
{\mu}_{\Lambda}=\mu_s,~~~
{\mu}_p+{\mu}_n=\mu_u+\mu_d,~~~
{\mu}_p-{\mu}_{\Sigma^+}=-{1\over 3}(\mu_d-\mu_s).
\eqno (12)$$

As we discussed in \cite{song9705}, an `one-parameter' scheme of the 
chiral quark model gives a good description to most existing spin and 
flavor observables. Here we will use the similar scheme, where the 
chiral parameters $a$, $\epsilon$ and $\zeta'$ are determined by 
$\Delta u-\Delta d=1.258$, $\bar d-\bar u=0.130$, and $\Delta s=-0.07$.
To predict the magnetic moments of the octet and decuplet baryons, we 
have adjusted $\mu_u$ as only one free parameter with two constraints 
of $\mu_s/\mu_d=2/3$ and $\mu_d/\mu_u=-0.45$. The numerical results for 
$k=1/3$ and $k=0$ are given in Table II, where the simple SU(6) quark
model (NQM) results are also listed. One can see that the agreement 
between the chiral quark model prediction and data is improved by 
including the OAM contributions ($k=1/3$). The flavor and spin contents, 
which are not directly related to the orbital motions, are listed in 
Table VII. 
\bigskip

\leftline{\quad\bf B. Decuplet baryons}
 
Similar to the octet baryons, the quark spin contents $\Delta q$ and 
the difference between the quark and antiquark orbital angular momenta 
$<L_z>_{u,d,s}-<L_z>_{\bar u,\bar d,\bar s}$ in the decuplet baryons 
are given in Table III. Using (1b) and Table III, the decuplet magnetic
moments are obtained. The numerical results and comparison with the 
data are listed in Table IV. 

It is easy to verify that the following {\it equal spacing} rules hold
for the decuplet baryons
$$\mu_{\Delta^{++}}-\mu_{\Delta^{+}}=
\mu_{\Delta^{+}}-\mu_{\Delta^{0}}=
\mu_{\Delta^{0}}-\mu_{\Delta^{-}}\equiv\delta_1
\eqno (13a)$$
$$\mu_{\Sigma^{*+}}-\mu_{\Sigma^{*0}}=
\mu_{\Sigma^{*0}}-\mu_{\Sigma^{*-}}=
\mu_{\Xi^{*0}}-\mu_{\Xi^{*-}}=\delta_1(\mu_u-\mu_d)
\eqno (13b)$$
$$
\mu_{\Delta^{+}}-\mu_{\Sigma^{*+}}=
\mu_{\Delta^{0}}-\mu_{\Sigma^{*0}}=
\mu_{\Delta^{-}}-\mu_{\Sigma^{*-}}=$$
$$=\mu_{\Sigma^{*0}}-\mu_{\Xi^{*0}}=\\
\mu_{\Sigma^{*-}}-\mu_{\Xi^{*-}}=
\mu_{\Xi^{*-}}-\mu_{\Omega^{*-}}=\delta_3(\mu_d-\mu_s)
\eqno (13c)$$
where $\delta_{1,3}$ are given in (11a) and (11c).
In the limit $a\to 0$, $\delta_{1,3}\to 1$, the simple quark model
results follow
$$\mu_{\Delta^{++}}=3\mu_u,~~
\mu_{\Delta^{+}}=2\mu_u+\mu_d,~~
\mu_{\Delta^{0}}=\mu_u+2\mu_d,~~
\mu_{\Delta^{-}}=3\mu_d,
\eqno (14a)$$
$$
\mu_{\Sigma^{*+}}=2\mu_u+\mu_s,~~
\mu_{\Sigma^{*0}}=\mu_u+\mu_d+\mu_s,~~
\mu_{\Sigma^{*-}}=\mu_u+2\mu_s,
\eqno (14b)$$
$$
\mu_{\Xi^{*0}}=\mu_u+2\mu_s,~~
\mu_{\Xi^{*-}}=\mu_d+2\mu_s,
\eqno (14c)$$
$$
\mu_{\Omega^{*-}}=3\mu_s.
\eqno (14d)$$
Since only two data, $\mu_{\Delta^{++}}$ and $\mu_{\Omega^-}$, are 
available, and theoretical predictions given by different models 
are quite similar as shown in Table V, hence one cannot make definite
conclusion on decuplet baryon magnetic moments.
 
\bigskip

\leftline{\bf IV. Quark Gluon Mixing (model II).}

In \cite{lipkin}, Lipkin suggested a hybrid model with the quark 
gluon mixing model of the nucleon. The proton is described by 
$$|p,J=J_z=1/2>={\rm cos}\theta |(3q)^{(0)}_{J=J_z=1/2}>
+{\rm sin}\theta
|[(3q)^{(8)}_{J_{3q}=1/2}\otimes(G)^{(8)}_{J_G=1}]^{(0)}_{J=J_z=1/2}>
\eqno (15)$$
where $\theta$ is the mixing angle, $|(3q)^{(0)}_{J=J_z=1/2}>$ is the 
spin-up ground state $SU(3)_f\otimes SU(2)_s$ proton wave function, in
which the three valence quarks are coupled to {\it a color singlet} (see 
the notation (0) superscripted in the first term on the right-hand side 
of (15)) with the total angular momentum $J=J_z=1/2$. The wave function 
$|[(3q)^{(8)}_{J_{3q}=1/2}\otimes(G)^{(8)}_{J_G=1}]^{(0)}_{J=J_z=1/2}>$
is also a spin up ground state proton wave function, but consists
of three valence quarks and a gluon. Since the gluon is a color-octet 
object, the three valence quarks in the $(3q+G)$ bound state must be
coupled to {\it a color-octet} (shown by superscript notation (8) in 
the second term on the right-hand side of (15)). In addition, the three 
quarks must be coupled to the total angular momentum $J_{3q}=1/2$. They 
coupled with a gluon (spin 1, color octet) to make a color singlet with 
total angular momentum $J_z=1/2$. 

We now apply the chiral dynamical mechanism to the model wave function
(15). For the first term in Eq.(15), all discussions given in the last
section can be used. However, they should be modified for the second 
term, i.e. the $3q+G$ piece. The valence quark numbers given in (8) 
for the standard $SU(3)\otimes SU(2)$ valence quark structure can no 
longer be used for the $3q+G$ piece, 
$|[(3q)^{(8)}_{J_{3q}=1/2}\otimes(G)^{(8)}_{J_G=1}]^{(0)}_{J=J_z=1/2}>$.
Instead, the valence quark numbers are
$$n_p^{3q+G,(v)}(u_{\up})={8\over 9}~,~~~
n_p^{3q+G,(v)}(u_{\dw})={{10}\over 9}~,~~~
n_p^{3q+G,(v)}(d_{\up})={4\over 9}~,~~~
n_p^{3q+G,(v)}(d_{\dw})={5\over 9}~,
\eqno (16)$$
where $n_p^{3q+G,(v)}(s_{\up,\dw})=0$ still hold for the $3q+G$ piece 
in the proton.

Assuming the mechanism of chiral fluctuation and the strength of these
fluctuations are the {\it same} as before, i.e. the chiral splitting 
processes {\it do not depend on} whether the three quarks coupled to a
gluon or not. Under this approximation, the spin-up and spin-down quark 
(antiquark) contents in the proton can be written as 
$$n_p(q'_{\up,\dw}, {\rm or}\
{\bar q'}_{\up,\dw})={\rm cos}^2\theta\cdot n_p^{(3q)}(q'_{\up,\dw}, 
{\rm or}\ {\bar q'}_{\up,\dw}) 
+{\rm sin}^2\theta\cdot n_p^{(3q+G)}(q'_{\up,\dw}, {\rm or}\ {\bar
q'}_{\up,\dw}) 
\eqno (17)$$
where
$$n_p^{(3q)}(q'_{\up,\dw}, {\rm or}\ {\bar q'}_{\up,\dw})= 
\sum\limits_{q=u,d}\sum\limits_{h=\up,\dw}
n_p^{(3q),(v)}(q_h)P_{q_h}(q'_{\up,\dw}, {\rm or}\ {\bar q'}_{\up,\dw})
\eqno (18a)$$
$$n_p^{(3q+G)}(q'_{\up,\dw}, {\rm or}\ {\bar q'}_{\up,\dw})= 
\sum\limits_{q=u,d}\sum\limits_{h=\up,\dw}
n_p^{(3q+G),(v)}(q_h)P_{q_h}(q'_{\up,\dw}, {\rm or}\ {\bar q'}_{\up,\dw})
\eqno (18b)$$
Eq.(18a) is the same as that given in (5), hence the results given in 
Sec. III can be directly used here. The only difference is that the first 
term on the right-hand side (RHS) of Eq.(17) has a factor cos$^2\theta$. 

For the second term in Eq.(17), one needs to calculate 
$n_p^{(3q+G)}(q'_{\up,\dw}, {\rm or}\ {\bar q'}_{\up,\dw})$ via
Eq.(18b), where the valence quark numbers $n_p^{(3q+G),(v)}(q_{\up,\dw})$ 
are given in (16). Using (6) and (7), one obtains the spin and orbital
contents for the $3q+G$ piece of the proton. Similar results can be 
obtained for the neutron and other octet baryons. All results are listed
in Table V. Two remarks on the nucleon sector are in order. 

(1) The total spin contents for $3q+G$ piece is
$$(\Delta\Sigma)_p^{(3q+G)}=-{1\over 3}+{{2a}\over 3}\xi_1
=-{1\over 3}[1-2a\xi_1]=-{1\over 3}(\Delta\Sigma)_p^{(3q)}
\eqno (19a)$$
and the total orbital angular momenta of quarks and antiquarks are
$$<L_z>_{q+\bar q}^{(3q+G)}=-ak\xi_1=-{1\over 3}<L_z>_{q+\bar q}^{(3q)}
\eqno (19b)$$
Eq. (19a) shows that total quark spin in the $3q+G$ state is one third 
of that in the $3q$ state. Using (17), one has
$$(\Delta\Sigma)_p^{[(3q)+(3q+G)]}=({\rm cos}^2\theta-{1\over 3}
{\rm sin}^2\theta)(\Delta\Sigma)_p^{(3q)}
\eqno (20a)$$
and
$$<L_z>_p^{[(3q)+(3q+G)]}=({\rm cos}^2\theta-{1\over 3}
{\rm sin}^2\theta)<L_z>_p^{(3q)}.
\eqno (20b)$$
Hence we have
$$<J_z>_{p,(q+\bar q)}^{[(3q)+(3q+G)]}=
({\rm cos}^2\theta-{1\over 3}{\rm sin}^2\theta)[{1\over 2}-(1-3k)a\xi_1],
\eqno (21)$$
where $<J_z>_p^{[(3q)+(3q+G)]}$ is the total angular momentum carried 
by the quarks and antiquarks in the gluon mixing model. 

Taking $\theta\to 0$, $<J_z>_G=0$, one obtains the chiral quark model
result without the gluon mixing effect (see Eq.(26a) in \cite{song9708}). 
For the limit $a\to 0$, (21) reduces to the result given in \cite{lipkin}.
It is easy to see from (20a) that since the total quark and antiquark spin 
in the proton has already been reduced by the chiral fluctuation mechanism 
(${1\over 2}\Delta\Sigma_p^{(3q)}={1\over 2}-a\xi_1<{1\over 2}$), to 
fit the smallness of $\Delta\Sigma$ indicated by DIS data, one does 
not necessarily resort to a large mixing angle as shown in \cite{lipkin}.

(2) From (21) and the proton spin sum rule 
$${1\over 2}\Delta\Sigma+<L_z>_{q+\bar q}+<J_z>_G={1\over 2}
\eqno (22a)$$
one obtains
$$({\rm cos}^2\theta-{1\over 3}{\rm sin}^2\theta)[{1\over 2}-(1-3k)a\xi_1]
+<J_z>_G={1\over 2}
\eqno (22b)$$
in the gluon mixing model. If we assume that the induced OAM arising 
from the chiral splittings is entirely and equally shared among quarks 
and antiquarks, and not shared by the gluons, then $k=1/3$. From (22), 
one has $<J_z>_G={2\over 3}$sin$^2\theta$. Assuming the gluon angular 
momentum is about $0.20\pm 0.10$ \cite{baji}, then sin$^2\theta\simeq
0.30\pm 0.15$, which implies there is a large gluon mixing in the proton.
However, if the gluon also shares the induced OAM, then $k<1/3$, and 
$$<J_z>_G={2\over 3}{\rm sin}^2\theta+(1-3k)a\xi_1({\rm cos}^2\theta
-{1\over 3}{\rm sin}^2\theta).
\eqno (23)$$
Assuming $<J_z>_G\simeq 0.15$, one obtains sin$^2\theta\simeq 0.05$ 
for $k\simeq 1/5$ and sin$^2\theta\simeq 0.13$ for $k\simeq 1/4$. 
If $<J_z>_G\simeq 0.20$, one obtains sin$^2\theta\simeq 0.15$ for 
$k\simeq 1/5$ and sin$^2\theta\simeq 0.21$ for $k\simeq 1/4$. 

To maintain the consistency, we have used the same constraints of
$\mu_s/\mu_d=2/3$ and $\mu_d=-0.45\mu_u$, and adjusted $\mu_u$ as a 
free parameter. For the gluon mixing, we fix $<J_z>_G=0.15$, and 
choose $k$ as adjusted parameter, then the mixing angle $\theta$ 
is not independent and determined by Eq.(23). The magnetic moments 
of the octet baryons in the hybrid gluon mixing model (model II) 
are given in Table VI. The corresponding predictions on the spin and 
flavor observables in the nucleon for both model I and model II are 
listed in Table VII. 

\bigskip
\leftline{\bf Summary}

(1) In the model I, the agreement between the magnetic moments of the 
octet baryons and data is improved by including the OAM contributions 
(see Table II). 

(2) The numerical result of baryon magnetic moments in the model II is 
almost the same as that given in the model I. For the sake of simplicity, 
the gluon mixing angle has been assumed to be universal (i.e. only {\it 
one} mixing angle for all octet baryons). If we would have been 
introduced three different mixing angles for N, $\Sigma$ and $\Xi$ 
isomultiplets as done in \cite{los} for $k=0$ case, where the OAM 
contributions were not included, we would have a better agreement with 
data.

(3) For the decuplet baryons, the agreement with two existing data
($\Delta^{++}$ and $\Omega^-$) looks very good for both $k=0$ and 
$k=1/3$ (Table IV). Since there is no significant difference between
the predictions for $k=1/3$ and $k=0$, to test the OAM effects, more
precise data in the decuplet sector is needed. Note that the same 
parameter $\mu_u$ and same constraints $\mu_s/\mu_d=2/3$ and 
$\mu_d=-0.45\mu_u$ are used as in the octet sector. 

(4) One can see from Table VII that the flavor and spin fractions 
given in both models I and II are in good agreement with data.
It should be noted that compared to the NMC data ($\bar d-\bar u=
0.147\pm 0.039$) \cite{nmc} and NA51 data ($[\bar u(x)/\bar d(x)]_
{x=0.18}=0.51\pm 0.06$) \cite{na51}, a smaller value of $\bar d-\bar u$ 
and a larger value of $\bar u(x)/\bar d(x)$ (or lower data points of 
$\bar d(x)/\bar u(x)$) have been reported \cite{e866}. All these 
data are quoted in Table VII. For the comparison of the chiral quark
model predictions on the flavor and spin observables with data, see 
discussion given in \cite{song9705}.

To summary, The symmetry breaking chiral quark model have been quite
successful in explaining many puzzles of the nucleon structure. By 
including the orbital angular momentum contributions, a better 
agreement with data for the baryon magnetic moments are also obtained. 

\bigskip

\leftline{\bf Acknowledgments}

The author thanks J. S. McCarthy and H. J. Weber for their useful
comments, and J. C. Peng for providing the new E866 data. The author 
also thanks J. Franklin for his useful comments on Eqs.(9a)-(9d), and
clarifying the relation between Eq.(9c) given in this paper and Eq.(16)
given in \cite{frank2}. This work was supported in part by the U.S. DOE
Grant No. DE-FG02-96ER-40950, the Institute of Nuclear and Particle 
Physics, University of Virginia, and the Commonwealth of Virginia.

\begin{table}[ht]
\begin{center}
\caption{The quark spin and orbital angular momenta in the
octet baryons in model {\bf I}, where $\xi_1=1+\epsilon+f$ 
and $\xi_2=2\epsilon+f_s$.}
\begin{tabular}{cccc} 
 Baryon &$\Delta u^{B}$ & $\Delta d^{B}$& $\Delta s^{B}$\\
\hline 
p & ${4\over 3}-{a\over 3}(8\xi_1-4\epsilon-5)$& 
$-{1\over 3}-{a\over 3}(-2\xi_1+\epsilon+5)$ & $-a\epsilon$\\
$\Sigma^+$ & ${4\over 3}-{a\over 3}(8\xi_1-5\epsilon-4)$& 
$-{a\over 3}(4-\epsilon)$ & $-{1\over 3}-{{2a}\over
3}(-\xi_2+3\epsilon)$\\
$\Sigma^0$ & ${2\over 3}-{a\over 3}(4\xi_1-3\epsilon)$& 
${2\over 3}-{a\over 3}(4\xi_1-3\epsilon)$ & $-{1\over 3}
-{{2a}\over 3}(-\xi_2+3\epsilon)$\\
$\Lambda^0$ & $-a\epsilon$&  $-a\epsilon$& 
$1-2a(\xi_2-\epsilon)$\\
$\Xi^0$ & $-{1\over 3}-{a\over 3}(-2\xi_1+5\epsilon+1)$& 
$-{a\over 3}(4\epsilon-1)$ & ${4\over 3}-{a\over
3}(8\xi_2-9\epsilon)$\\
\hline 
&$<L_z>_{u-\bar u}^B$& $<L_z>_{d-\bar d}^B$& $<L_z>_{s-\bar s}^B$\\
\hline 
p & ${4\over 3}ka\xi_1$& 
$-{1\over 3}ka\xi_1$& 0\\
$\Sigma^+$ & ${4\over 3}ka\xi_1$& 0&
$-{1\over 3}ka\xi_2$\\
$\Sigma^0$ & ${2\over 3}ka\xi_1$& ${2\over 3}ka\xi_1$&
$-{1\over 3}ka\xi_2$\\
$\Lambda^0$ & 0& 0&
$ka\xi_2$\\
$\Xi^0$ & $-{1\over 3}ka\xi_1$& 0&
${4\over 3}ka\xi_2$\\
\end{tabular}
\end{center}
\end{table}

\begin{table}[ht]
\begin{center}
\caption{Comparison of our predictions with data for the octet baryon
magnetic moments in the model {\bf I}. The naive quark model (NQM)
results are also listed.} 
\begin{tabular}{ccccc} 
 Baryon& data & k=1/3  & k=0  & NQM \\
\hline 
p          &    $ 2.79\pm 0.00$&   2.68 &  2.63 &  2.87\\
n          &    $-1.91\pm 0.00$& $-1.91$&$-1.91$&$-1.91$\\
$\Sigma^+$ &    $ 2.46\pm 0.01$&   2.56 &  2.53 &  2.62\\
$\Sigma^-$ &    $-1.16\pm 0.03$& $-1.11$&$-1.11$&$-1.20$ \\
$\Lambda^0$&    $-0.61\pm 0.00$& $-0.63$&$-0.67$&$-0.63$ \\ 
$\Xi^0$    &    $-1.25\pm 0.01$& $-1.43$&$-1.47$&$-1.49$\\
$\Xi^-$    &    $-0.65\pm 0.00$& $-0.52$&$-0.56$&$-0.53$\\
$\Sigma^0$ &    $-$            &   0.73 &  0.71 &  0.71\\
\hline
&&$\mu_s={2\over 3}\mu_d$& $\mu_s={2\over 3}\mu_d$& $\mu_s={2\over
3}\mu_d$\\
& &$\mu_d=-0.45\mu_u$ &$\mu_d=-0.45\mu_u$ & $\mu_d=-0.50\mu_u$ \\
& &$\mu_u=2.22\mu_N$ &$\mu_u=2.49\mu_N$& $\mu_u=1.91\mu_N$\\
\end{tabular}
\end{center}
\end{table}

\begin{table}[ht]
\begin{center}
\caption{The quark spin and orbital angular momenta in the
decuplet baryons in model {\bf I}.}
\begin{tabular}{cccc} 
 Baryon &$\Delta u^{B^*}$ & $\Delta d^{B^*}$& $\Delta s^{B^*}$\\
\hline 
$\Delta^{++}$ & $3-3a(2\xi_1-\epsilon-1)$& $-3a$ & $-3a\epsilon$\\
$\Delta^{+}$ & $2-a(4\xi_1-2\epsilon-1)$& $1-a(2\xi_1-\epsilon+1)$&
$-3a\epsilon$\\
$\Sigma^{*0}$ & $1-2a\xi_1$& $1-2a\xi_1$& 
$1-2a\xi_2$\\
$\Sigma^{*+}$ & $2-a(4\xi_1-\epsilon-2)$& $-a(\epsilon+2)$ &$1-2a\xi_2$\\
$\Xi^{*0}$ & $1-a(2\xi_1+\epsilon-1)$& $-a(2\epsilon+1)$ &
$2-a(4\xi_2-3\epsilon)$\\
$\Omega^{-}$ & $-3a\epsilon$& $-3a\epsilon$ &
$3-6a(\xi_2-\epsilon)$\\
\hline
 Baryon &$<L_z>_{u-\bar u}^{B^*}$ & $<L_z>_{d-\bar d}^{B^*}$& 
$<L_z>_{s-\bar s}^{B^*}$\\
\hline 
$\Delta^{++}$ & $3ka\xi_1$& 0 & 0\\
$\Delta^{+}$ & $2ka\xi_1$& $ka\xi_1$&0\\
$\Sigma^{+}$ & $2ka\xi_1$&0& $ka\xi_2$\\
$\Sigma^{*0}$ & $ka\xi_1$& $ka\xi_1$& $ka\xi_2$\\
$\Xi^{*0}$ & $ka\xi_1$& 0& $2ka\xi_2$\\
$\Omega^{-}$ & 0&0& $3ka\xi_2$\\
\end{tabular}
\end{center}
\end{table}

\begin{table}[ht]
\begin{center}
\caption{Comparison of our predictions with data for the decuplet 
baryon magnetic moments in the model {\bf I}. The naive quark model 
(NQM) results are also listed.} 
\begin{tabular}{ccccc} 
 Baryon & data  & k=1/3  & k=0  & NQM \\
\hline 
$\Delta^{++}$ & $4.52\pm 0.50\pm 0.45^a$&  5.30 &  5.17  &  5.73 \\
 & $3.7~<~\mu_{\Delta^{++}}~<~7.5^b$ &   &   &\\
$\Delta^{+}$  & $-$                   &  2.54 &  2.45  &  2.87 \\
$\Delta^{0}$  & $-$                   &$-0.22$&$-0.27$ &  0.00 \\
$\Delta^{-}$  & $-$                   &$-2.98$&$-2.99$ &$-2.87$\\ 
$\Sigma^{*+}$ & $-$                   &  2.94 &  2.78  &  3.18 \\
$\Sigma^{*0}$ & $-$                   &  0.18 &  0.06  &  0.32 \\
$\Sigma^{*-}$ & $-$                   &$-2.58$&$-2.66$ &$-2.55$\\
$\Xi^{*0}$    & $-$                   &  0.49 &  0.39  &  0.64 \\
$\Xi^{*-}$    & $-$                   &$-2.27$&$-2.33$ &$-2.23$\\
$\Omega^{-}$  &$-2.02\pm 0.05^b$&$-1.91$&$-2.00$ &$-1.91$\\
 &$-2.024\pm 0.056^c$& & &\\
 &$-1.94\pm 0.17\pm 0.14^d$& & &\\
\hline
&a $-$ \cite{bos91}, b $-$ \cite{pdg96}&$\mu_s={2\over 3}\mu_d$& 
$\mu_s={2\over 3}\mu_d$& $\mu_s={2\over 3}\mu_d$\\
&c $-$ \cite{wal95}, d $-$ \cite{diehl91} &$\mu_d=-0.45\mu_u$ &
$\mu_d=-0.45\mu_u$ & $\mu_d=-0.50\mu_u$ \\
& &$\mu_u=2.22\mu_N$ &$\mu_u=2.49\mu_N$& $\mu_u=1.91\mu_N$\\
\end{tabular}
\end{center}
\end{table}

\begin{table}[ht]
\begin{center}
\caption{The quark spin and orbital angular momenta in the octet baryons 
for $(3q+G)$ piece in the model $\bf II$.} 
\begin{tabular}{cccc} 
 Baryon &$\Delta u^{B}$ & $\Delta d^{B}$& $\Delta s^{B}$\\
\hline 
p & $-{2\over 9}+{a\over 9}(4\xi_1-2\epsilon-1)$& 
$-{1\over 9}+{a\over 9}(2\xi_1-\epsilon+1)$ & ${a\over 3}\epsilon$\\
$\Sigma^+$ & $-{2\over 9}+{a\over 9}(4\xi_1-\epsilon-2)$& 
${a\over 9}(\epsilon+2)$ & $-{1\over 9}+{{2a}\over 9}\xi_2$\\
$\Sigma^0$ & $-{1\over 9}+{{2a}\over 9}\xi_1$& 
$-{1\over 9}+{{2a}\over 9}\xi_1$ & $-{1\over 9}+{{2a}\over 9}\xi_2$\\
$\Lambda^0$ & ${1\over 3}a\epsilon$&  ${1\over 3}a\epsilon$& 
$-{1\over 3}+{{2a}\over 3}(\xi_2-\epsilon)$\\
$\Xi^0$ & $-{1\over 9}+{{a}\over 9}(2\xi_1+\epsilon-1)$& 
${a\over 9}(2\epsilon+1)$ & $-{2\over 9}+{a\over 9}(4\xi_2-3\epsilon)$\\
\hline 
&$<L_z>_{u-\bar u}^B$& $<L_z>_{d-\bar d}^B$& $<L_z>_{s-\bar s}^B$\\
\hline 
p & $-{2\over 9}ka\xi_1$& 
$-{1\over 9}ka\xi_1$& 0\\
$\Sigma^+$ & $-{2\over 9}ka\xi_1$& 0&
$-{1\over 9}ka\xi_2$\\
$\Sigma^0$ & $-{1\over 3}ka\xi_1$& $-{1\over 9}ka\xi_1$&
$-{1\over 9}ka\xi_2$\\
$\Lambda^0$ & 0& 0&
$-{1\over 3}ka\xi_2$\\
$\Xi^0$ & $-{1\over 9}ka\xi_1$& 0&
$-{2\over 9}ka\xi_2$\\
\end{tabular}
\end{center}
\end{table}

\begin{table}[ht]
\begin{center}
\caption{Comparison of our predictions with data for the octet baryon
magnetic moments in the model {\bf II}, where $<J_z>_G=0.15$ and 
sin$^2\theta$ is assumed to be universal for all octet baryons.
The naive quark model (NQM) results are the same as given in Table II.}
\begin{tabular}{ccccc} 
 Baryon& data & k=1/5  &k=1/4 & NQM   \\
  & & (sin$^2\theta$=0.05)&(sin$^2\theta$=0.13)&\\
\hline 
p          & 2.79$\pm$ 0.00 &   2.68 &  2.65 &  2.87\\
n          & $-1.91\pm 0.00$& $-1.91$&$-1.91$&$-1.91$\\
$\Sigma^+$ & $2.46\pm 0.01$ &   2.55 &  2.54 &  2.62\\
$\Sigma^-$ & $-1.16\pm 0.03$& $-1.09$&$-1.06$&$-1.20$ \\
$\Lambda^0$& $-0.61\pm 0.00$& $-0.63$&$-0.62$&$-0.63$ \\ 
$\Xi^0$    & $-1.25\pm 0.01$& $-1.44$&$-1.47$&$-1.49$\\
$\Xi^-$    & $-0.65\pm 0.00$& $-0.52$&$-0.51$&$-0.53$\\
$\Sigma^0$ &    $-$            &   0.72 &  0.73 &  0.71\\
\hline
&&$\mu_s={2\over 3}\mu_d$& $\mu_s={2\over 3}\mu_d$& $\mu_s={2\over
3}\mu_d$\\
& &$\mu_d=-0.45\mu_u$ &$\mu_d=-0.45\mu_u$ & $\mu_d=-0.50\mu_u$ \\
& &$\mu_u=2.46\mu_N$ &$\mu_u=2.69\mu_N$& $\mu_u=1.91\mu_N$\\
\end{tabular}
\end{center}
\end{table}

\begin{table}[ht]
\begin{center}
\caption{Quark spin and flavor observables in the proton (the values 
with * are inputs) in models {\bf I} and {\bf II}.}
\begin{tabular}{ccccc} \hline
Quantity & Data& Model {\bf I} &Model {\bf II} &NQM\\
\hline 
$\bar d-\bar u$ & $0.147\pm 0.039^a$ & $0.130^*$ & $0.130^*$ &0\\
& $0.100\pm 0.018^b$ & & & \\
${{\bar u}/{\bar d}}$ & $[{{\bar u(x)}\over {\bar d(x)}}]_{x=0.18}=0.51\pm
0.06^c$& 0.68 & 0.68&  $-$\\
 & $[{{\bar u(x)}\over {\bar d(x)}}]_{0.1<x<0.2}=0.67\pm 0.06^b$& && \\
${{2\bar s}/{(\bar u+\bar d)}}$ & ${{<2x\bar s(x)>}\over {<x(\bar
u(x)+\bar d(x))>}}=0.477\pm 0.051^d$& 0.72 & 0.77&$-$\\
 ${{2\bar s}/{(u+d)}}$ & ${{<2x\bar s(x)>}\over
{<x(u(x)+d(x))>}}=0.099\pm0.009^d$ & 0.13 &0.14&0\\
 ${{\sum\bar q}/{\sum q}}$ & ${{\sum<x\bar q(x)>}\over{\sum<xq(x)>}}
=0.245\pm 0.005^d$ & 0.24 & 0.24&0\\
 $f_s$ & $0.10\pm 0.06^e$ & 0.10 & 0.11  & 0\\
       & $0.15\pm 0.03^f$ &      &   &    \\
       & ${{<2x\bar s(x)>}\over {\sum<x(q(x)+\bar q(x))>}}
=0.076\pm 0.022^d$ & & &  \\
$f_3/f_8$ & $0.21\pm 0.05^g$ & 0.22 &0.22   &1/3\\
\hline 
$\Delta u$ & $0.85\pm 0.05^h$ & 0.86 &0.80& 4/3\\
$\Delta d$&$-0.41\pm 0.05^h$ &$-$0.40&$-$0.38&$-$1/3\\
$\Delta s$&$-0.07\pm 0.05^h$ &$-0.07^*$&$-0.07^*$&0\\
$\Delta\bar u$, $\Delta\bar d$ & $-0.02\pm 0.11^i$ &0&0&0 \\
$\Delta_3/\Delta_8$ &2.17$\pm 0.10^j$&2.12&2.11 & 5/3\\
\hline
&a $-$ \cite{nmc}, b $-$ \cite{e866}&&&\\
& c $-$ \cite{na51}, d $-$ \cite{ccfr95}&&&\\
&e $-$ \cite{gls91}, f $-$ \cite{dll95}&&&\\
& g $-$ \cite{cl1}, h $-$ \cite{smc97}&&&\\
&i $-$ \cite{smc96}, j $-$ \cite{pdg96}&&&\\
\end{tabular}
\end{center}
\end{table}


\bigskip
\bigskip
\bigskip

\begin{references}

\bibitem{emc} 
	J.~Ashman {\t et al.}, {\pl} {\bf B206}, 364 (1988); {\np}
{\bf B328}, 1 (1989).

\bibitem{smc}
        B.~Adeva {\it et al.}, {\pl} {\bf B302}, 533 (1993);
D.~Adams {\it et al.}, {\pl} {\bf B329}, 399 (1994); 
(E) {\bf B339}, 332 (1994); {\bf B357}, 248 (1995).

\bibitem{slac}
        P.~L.~Anthony {\it et al.} {\prl} {\bf 71}, 959 (1993);
K.~Abe {\it et al.}, {\prl} {\bf 74}, 346 (1995); {\bf 75}, 25
(1995); {\bf 79}, 26 (1997).

\bibitem{hermes}
	K.~Ackerstaff, {\it et al.}, {\pl} {\bf B404}, 383 (1997).

\bibitem{sehgal}
	L.~M.~Sehgal, {\pr} {\bf D10}, 1663 (1974).

\bibitem{karl} 
	G.~Karl, {\pr} {\bf D45}, 247 (1992).

\bibitem{br} 
	J.~Bartelski and R.~Rodenberg, {\pr} {\bf D41}, 2800 (1990).

\bibitem{sg} 
	X.~Song and V. Gupta, Phys. Rev. $\bf D49$, 2211 (1994).

\bibitem{rat87}
	P.~G.~Ratcliffe, {\pl} {\bf B192}, 180 (1987).

\bibitem{bali}
	B.~A.~Li, Proceedings of 11th International Symposium on High
Energy Spin Physics, Bloomington, Indiana, USA, pp.802 (1994)

\bibitem{ji}
     	X.~Ji, J.~Tang and P.~Hoodbhoy, {\prl} {\bf 76}, 740 (1996).

\bibitem{cs97}
	M.~Casu and L.~M.~Sehgal, {\pr} {\bf D55}, 2644 (1997).

\bibitem{cl2} 
     	T.~P.~Cheng and L.-F.~Li, {\prl} {\bf 80}, 2789 (1998).

\bibitem{song9708} 
	X.~Song, hep-ph/9802206.

\bibitem{frank1}
	J.~Franklin, {\pr} {\bf 182}, 1607 (1969); {\pr} {\bf D20}, 1742
(1979).

\bibitem{frank2}
	J.~Franklin, {\pr} {\bf D29}, 2648 (1984).

\bibitem{bs90} 
	S.~J.~Brodsky, and I.~Schmidt, {\pl} {\bf B234}, 144 (1990).

\bibitem{smc96}
        B.~Adeva {\it et al.} {\prl} {\bf B369}, 93 (1996). 

\bibitem{wb} 
	H.~J.~Weber, and K.~Bodoor,  Int. J. Mod. Phys. {\bf E6},
693 (1997).

\bibitem{los} 
	J.~Linde, T.~Ohlsson, and H.~Snellman, {\pr} {\bf D57}, 452
(1998); {\pr} {\bf D57}, 5916 (1998).

\bibitem{pdg96}
	Particle Data Group, R.~M.~Barnett et al., {\pr} {\bf D54}, 1
(1996).

\bibitem{song9705} 
	X.~Song, Phys. Rev. {\bf D57}, 4114 (1998).

\bibitem{lipkin} 
	H.~J.~Lipkin, {\pl} {\bf B251}, 613 (1990).

\bibitem{baji} 
	I.~Balitsky and X. Ji, {\prl} {\bf 79}, 1225 (1997).

\bibitem{bos91} 
	A.~Bosshard {\it et al.,} {\pr} {\bf D44}, 1962 (1991).

\bibitem{wal95} 
	N.~B.~Wallace {\it et al.}, {\prl} {\bf 74}, 3732 (1995).

\bibitem{diehl91} 
	H.~T.~Diehl {\it et al.}, {\prl} {\bf 67}, 804 (1991).

\bibitem{nmc}
        P.~Amaudruz {\it et al.}, {\prl} {\bf 66}, 2712 (1991); 
M.~Arneodo {\it et al.}, {\pr} {\bf D50}, R1, (1994).

\bibitem{na51} 
        A.~Baldit {\it et al.}, NA51 Collaboration, {\pl} {\bf B332}, 
{244} (1994).

\bibitem{e866} 
        J.~C.~Peng {\it et al.}, E866 Collaboration, hep-ex/9804288.

\bibitem{ccfr95} 
        A.~O.~Bazarko, et. al, {\sl Z. Phys.} {\bf C65}, 189 (1995). 

\bibitem{gls91} 
        J.~Gasser, H.~Leutwyler, and M.~E.~Saino, {\sl Phys. Lett.}
{\bf B253}, 252 (1991).

\bibitem{dll95}
	S.~J.~Dong, J.-F.~Lagae, and K.~F.~Liu, {\prl}
{\bf 75}, 2096 (1995). 

\bibitem{cl1} 
     	T.~P.~Cheng and L.-F.~Li, {\prl} {\bf 74}, {2872} (1995).

\bibitem{smc97}
	D.~Adams, {\it et al.}, {\pr} {\bf D56}, 5330 (1997).





\end{references}
\end{document}